\documentclass{ws-p8-50x6-00}

\newcommand{\qsq}{\mbox{$Q^2$}}
\newcommand{\pom}{I\!\!P}
\newcommand{\reg}{{I\!\!R}}
\newcommand{\xpom}{x_{\pom}}

\newcommand{\bet} {$\beta$}

\newcommand{\ttt} {$t$}

\newcommand{\fdthree} {$F_2^{D(3)}$}
\newcommand{\fdthreef} {$F_2^{D(3)}  (Q^2, x_{I\!\!P} , \beta )$}
\newcommand{\fdtwo} {$F_2^{D(2)} (Q^2,\beta)$}
\newcommand{\ftwo} {$F_2 (Q^2,x)$}

\newcommand{\fnrs}{Chercheur Qualifi\'e FNRS}

\newcounter{fig:gluon}
\newcounter{fig:diag}
\newcounter{fig:jpsi}
\newcounter{fig:rho}
 
\def\Journal#1#2#3#4{{#1} {\bf #2} (#3) #4}

\def\NPB{{\em Nucl. Phys.} {\bf B}}
\def\PLB{{\em Phys. Lett.} {\bf B}}

\def\PRD{{\em Phys. Rev.} {\bf D}}
\def\ZPC{{\em Z. Phys.} {\bf C}}
\def\EPJ{{\em Eur. Phys. J.} {\bf C}}
 
\begin{document}

\title{Diffractive Physics at HERA}

\author{L. Favart}
\address{\fnrs \\
        Universit\'e Libre de Bruxelles, 
        1050 Brussels, Belgium \\
        {\it lfavart@ulb.ac.be}}

% typeset front matter
\maketitle

\abstracts{
 A great interest for diffraction has been generated these last years
by the HERA data. They give us the first opportunity to understand
high energy diffractive physics in terms of a fundamental theory, i.e.
QCD.
A review of the main results from H1 and ZEUS in this field is presented
and the relation with the proton structure function at small x-Bjorken 
values is discussed.
}

%%%%%%%%%%%%%%%%%%%%%%%%%%%%%%%%%%%%%%%%%%%%%%%%%%%%%%%%%%%%%%%%%%%%%%%%%
\section*{Introduction}
%%%%%%%%%%%%%%%%%%%%%%

Presently, one of the most important tasks in particle physics is the 
understanding of the strong force.
For this purpose, the Quantum Chromodynamics theory (QCD) 
seems to be the best candidate.
An important characteristic of this theory is that the 
coupling constant $\alpha_S$ is larger than 1 when the interaction 
between the different actors, quarks and gluons, does not take place at 
distances small enough.
This means that to be calculable within a 
perturbative approach, the interaction between these constituents
requires the presence in the process of a ``hard''
scale, i.e. a large transverse momentum or a large mass.
\\

The lepton beam of the high energy $e-p$ collider 
HERA (reaching an energy in the center of mass system $\sqrt s = 300$ GeV) 
is a prolific source of photons in a large virtuality range 
such that the study of $\gamma^* p$ interactions provides a completely new 
and deep insight into the QCD dynamics.
\\

A major discovery at HERA is the observation of the strong rise of 
the total cross section at high energy in the deep inelastic scattering 
(DIS) i.e.\ $\gamma^* p$ interactions with large $Q^2$ values
(\qsq\ being the negative of the squared four-momentum of the exchanged photon).
This is inconsistent with the case of the photoproduction ($Q^2 \simeq 0$),
which shows a soft dependence in the total hadronic 
energy, $W$, similar to the hadron-hadron interaction case
and well described by the Regge phenomenological theory.
\\

\hspace{-0.8cm}
\begin{minipage}{0.48\textwidth}
\stepcounter{figure}
 \setcounter{fig:gluon}{\value{figure}} 
HERA is thus a unique device to test QCD in the perturbative 
regime and to study the transition between perturbative and non-perturbative
domains. 
One of the remarkable successes of this theory 
is the correct prediction of the evolution of
the proton structure function with $Q^2$, for $Q^2 > 1$~GeV$^2$,
now measured with a very good acuracy\cite{f2}. 
This evolution allows the extraction of the gluon density in the 
proton, which is not directly measurable. 
The fast rise of the gluon density for decreasing $x$, shown 
in Fig.~\arabic{fig:gluon} reflects the fast rise of the cross section with
$W$.
\end{minipage}

\vspace{-6.50cm}
\hspace{0.45\textwidth}
\begin{minipage}{0.48\textwidth}
  \epsfig{file=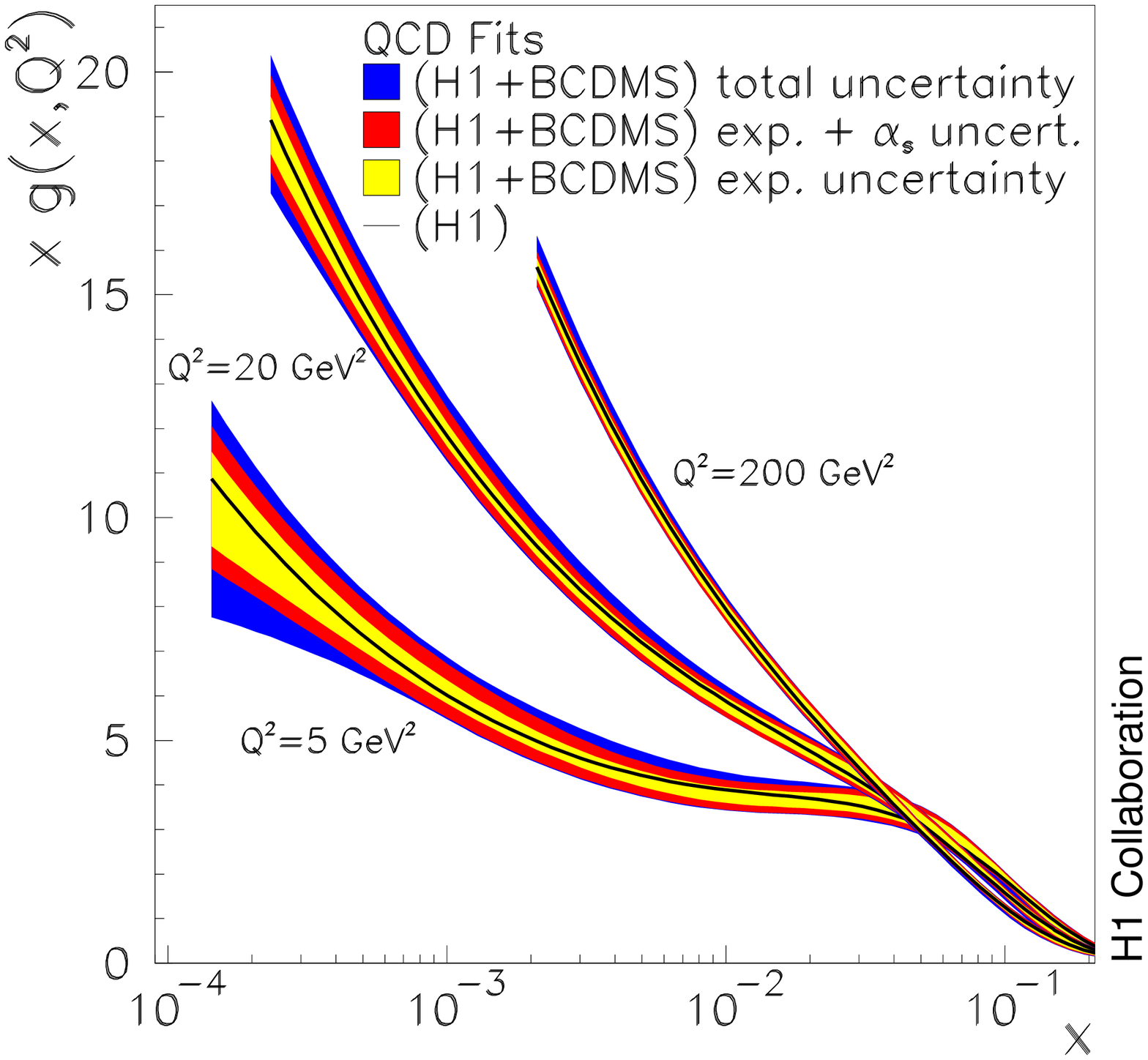,width=0.98\textwidth}
  \small{Figure~\arabic{fig:gluon}. The gluon density in the proton
  measured by H1 as a function of $x$, for different values of \qsq.}
  \\
\end{minipage}

The other main window opened at HERA for the understanding of the strong force
is the study of diffractive interaction.
Diffraction has been successfully described, 
already more than 30 years 
ago, via the introduction of an exchanged object carrying the vacuum quantum 
numbers, called the Pomeron ($\pom$).
Whilst Regge-based models give a unified description of all pre-HERA
diffractive data, this approach is not linked to an underlying
fundamental theory.
\\

\hspace{-8.0mm}
\begin{minipage}[h]{0.48\textwidth}
 \stepcounter{figure}
 \setcounter{fig:diag}{\value{figure}}
The second major result at HERA is the observation that $8 - 10 \%$ of
the events in the deep inelastic regime ($Q^2 > 1$~GeV$^2$) present
the charateristics of diffractively produced events\cite{diffr_1992}, i.e. a
large rapidity gap (LRG) without hadronic activity between the 
two hadronic low mass sub-systems, $X$ and $Y$, 
as illustrated in Fig.~\arabic{fig:diag}.The gaps being significantly
larger than implied by particle density
fluctuation during the hadroni-
\end{minipage}
\hspace{0.5 cm}
\begin{minipage}[h]{0.48\textwidth}
 \vspace{0.3cm}
 \epsfig{file=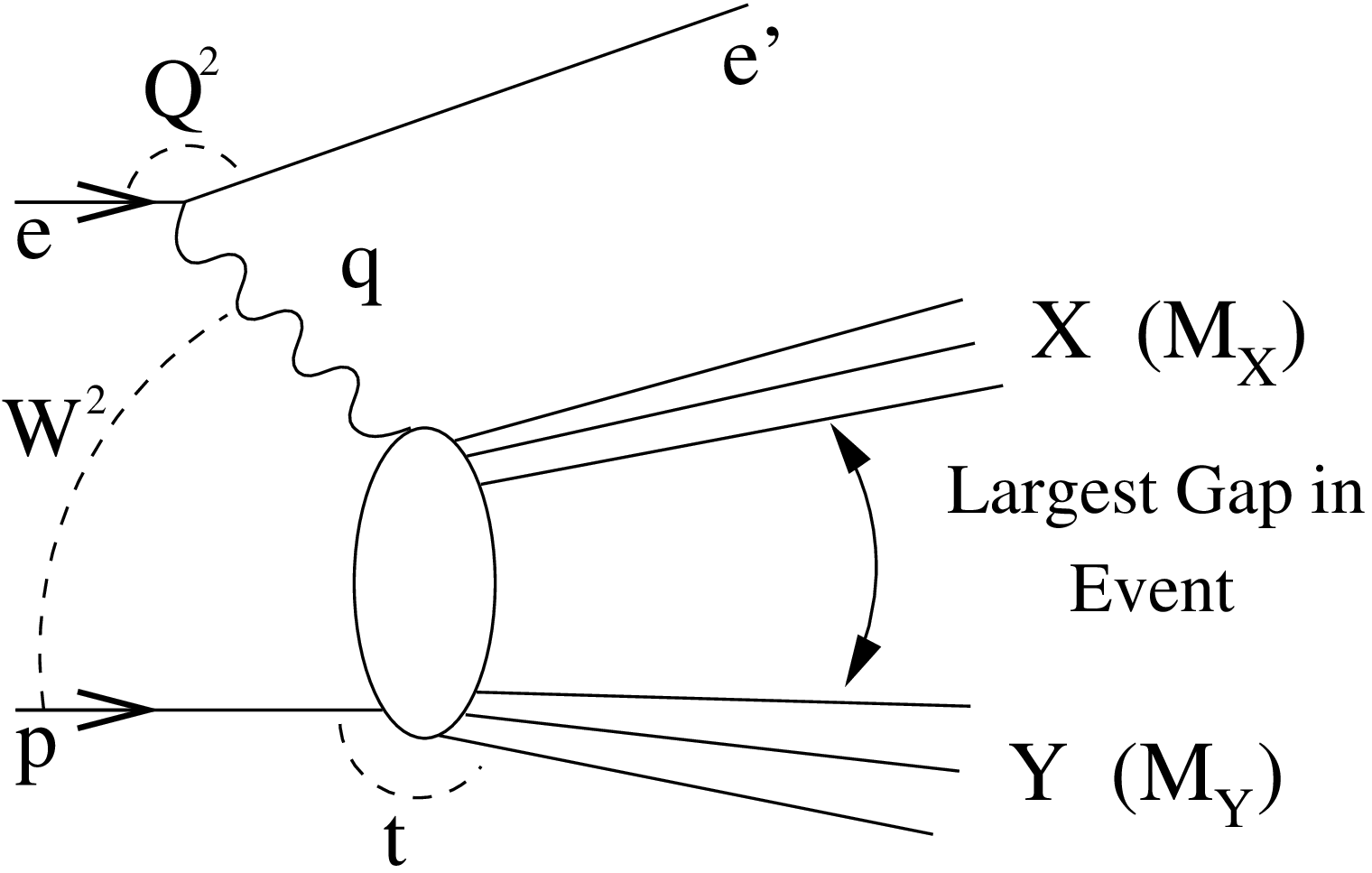,angle=270,width=0.98\textwidth}
 \small{Figure~\arabic{fig:diag}. Sketch of the diffractive $e-p$ interaction}
 \vspace*{2mm}
 \\
\end{minipage}
sation process, these events
are attributed to ``hard" diffraction, i.e. to the exchange of a 
colourless object at the proton vertex, for the first time observed in a
hard regime. HERA offers a unique possibility to study the nature of
diffraction and determine the Pomeron structure in terms of QCD.

%%%%%%%%%%%%%%%%%%%%%%%%%%%%%%%%%%%%%%%%%%%%%%%%%%%%%%%%%%%%%%%%%%%%%%%%%
\section{Inclusive DIS cross section and partonic structure of the Pomeron}
%%%%%%%%%%%%%%%%%%%%%%%%%%%%%%%%%%%%%%%%%%%%%%%%%%%%%%%%%%%%%%%%%%%%%%%

The hard diffractive cross section can be defined by four kinematic variables
conveniently chosen as  \qsq , $\xpom$ , \bet\ and \ttt, where 
\ttt \ is the squared
four-momentum transfer to the proton, and $\xpom$ \ and \bet\ are defined as
\begin{center}
\begin{equation}
   \xpom \simeq \frac{Q^2 + M_X^2}{Q^2 + W^2 }
   \qquad
    \beta \simeq \frac{Q^2 }{Q^2 + M_X^2} ; \label{eq:kin}
\end{equation}
\end{center}
$\xpom$ can be interpreted as the fraction of the 
proton momentum carried by the exchanged
Pomeron and \bet\  is the fraction of the exchanged momentum carried by
the quark struck by the photon.
These variables are related to the Bjorken $x$ scaling variable 
by the relation $x = \beta \cdot \xpom$.

Experimentally, the \ttt\ variable is usually not measured or is integrated over.
In analogy with non-diffractive DIS scattering, the measured cross section
is expressed in the form of a three-fold diffractive structure function
\fdthreef :
\begin{equation}
  \frac { {\rm d}^3 \sigma \ (e p \rightarrow e X Y) }
 { {\rm d}Q^2 \ {\rm d}\xpom \ {\rm d}\beta}
        = \frac {4 \pi \alpha^2} {\beta Q^4}
            \ (1 - y + \frac {y^2}{2} )
            \ F_2^{D(3)} (Q^2, \xpom , \beta) ,
                                            \label{eq:fdthreefull}
\end{equation}
where $y$ is the usual scaling variable, with
$  y \simeq W^2 / s$.
Conveniently, \fdthree \ is factorised in the form
$F_2^{D(3)} (Q^2, \xpom , \beta ) =
     f_{\pom/p} (\xpom ) \cdot F_2^D (Q^2, \beta)$,
assuming that the $\pom$ flux $ f_{\pom/p} (\xpom)$ is independent
of the  $\pom$ structure $F_2^D (Q^2, \beta)$, 
by analogy with the hadron structure functions, \bet\ playing the role 
of Bjorken $x$. The  $\pom$ flux is parametrized in a Regge inspired
form. Including the reggeon ($\reg$) trajectory
in addition to the Pomeron, 
a good description of the data is obtained throughout
the measured kinematic domain ($6.5<Q^2<800$ GeV$^2$ $\xpom < 0.05$
and $0.01<\beta<0.9$)\cite{f2d}.
\\

\begin{figure}[htb] \unitlength 1mm
 %\vspace{-4.2cm}
 \epsfig{file=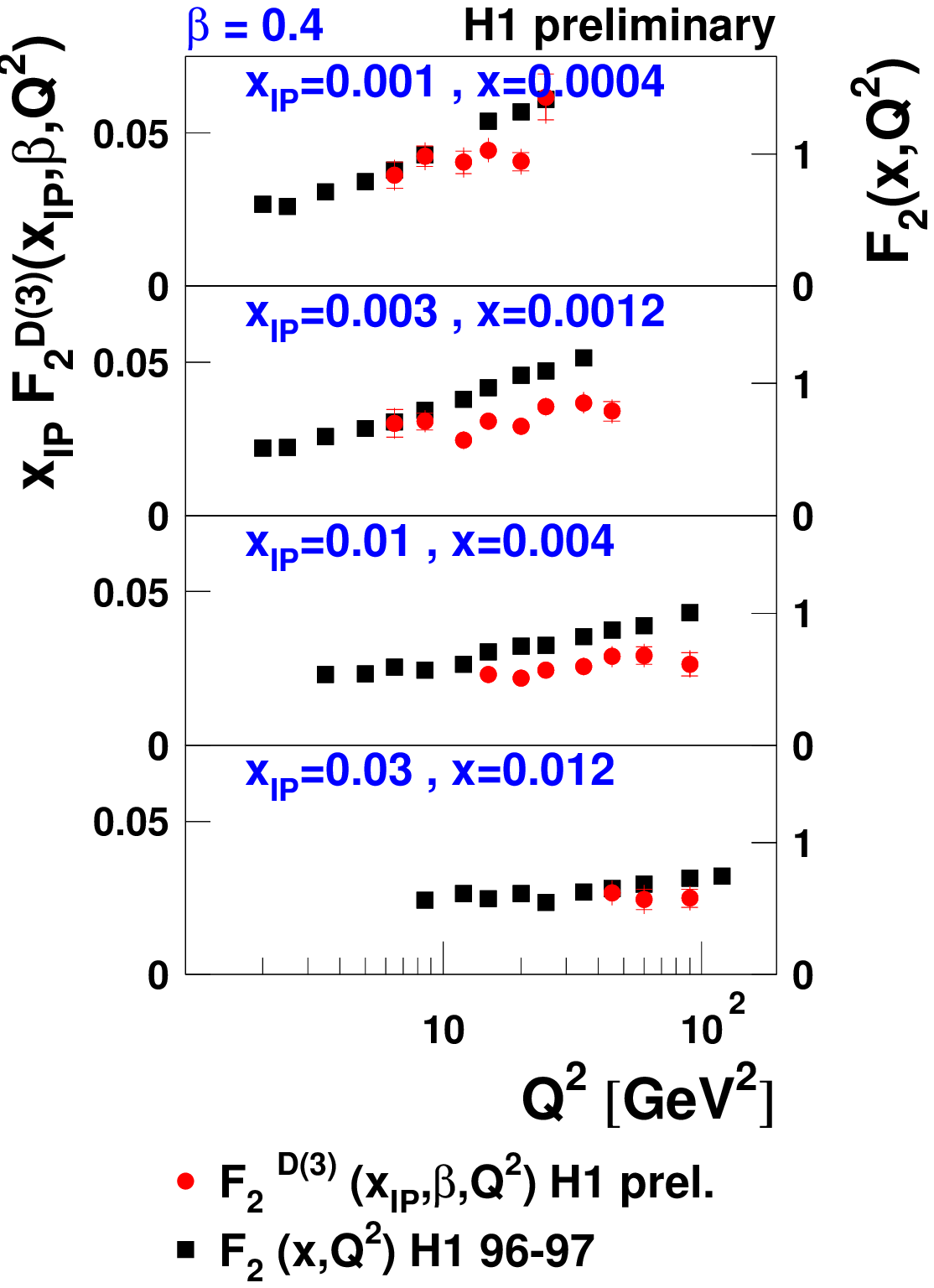,width=0.49\textwidth}
 \epsfig{file=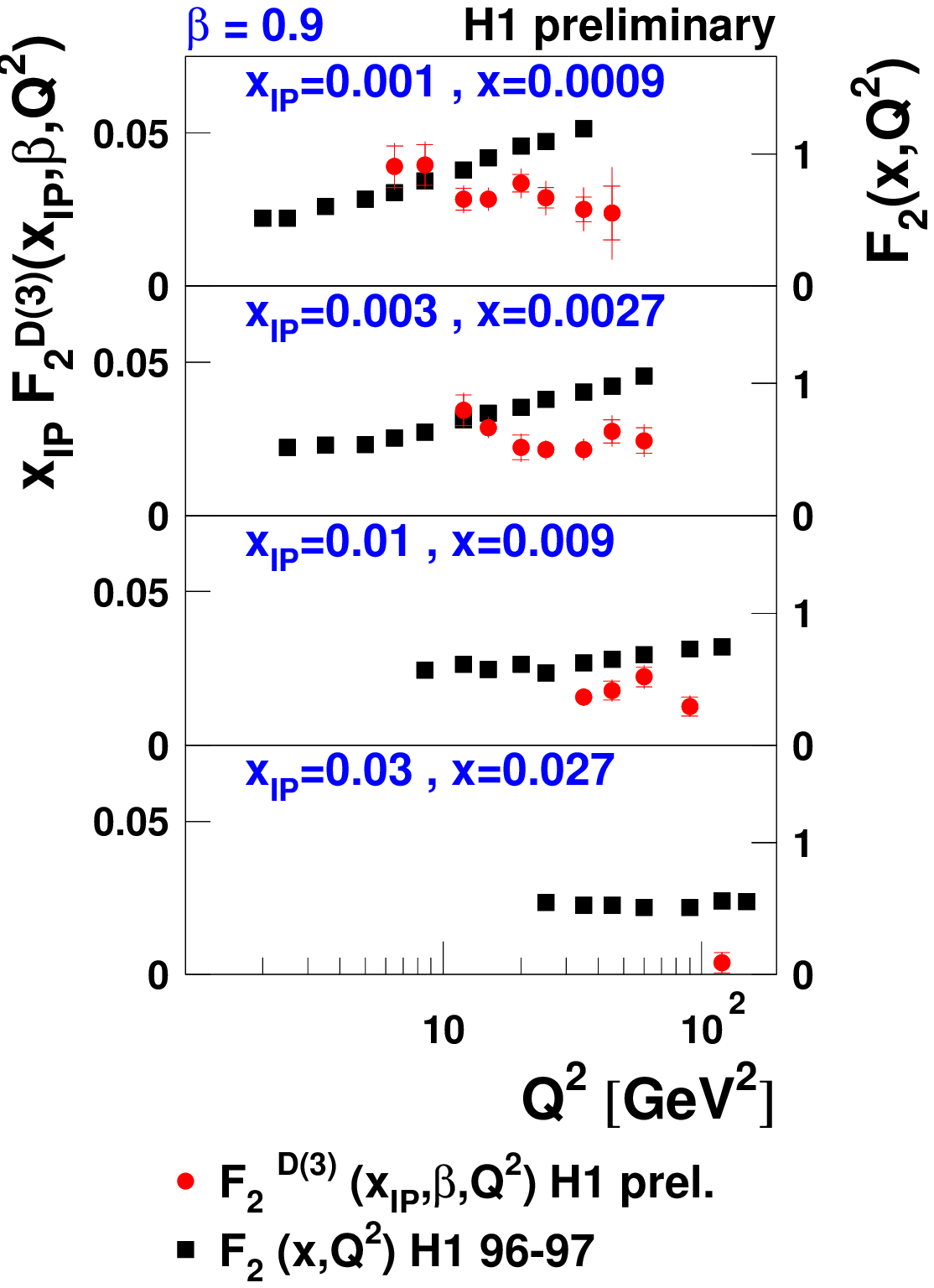,width=0.49\textwidth}
 \caption{
The \fdthree diffractive structure function measurement as a function 
of $Q^2$ for different values of $\xpom$ in two bins in \bet. The
inclusive proton structure function,
\ftwo\, is overlayed for the same \qsq\ values and 
$ x = \beta \cdot \xpom$.
}
\label{fig:f2d}
\end{figure}

\begin{figure}[htb] \unitlength 1mm
 %\vspace{-4.2cm}
   \epsfig{file=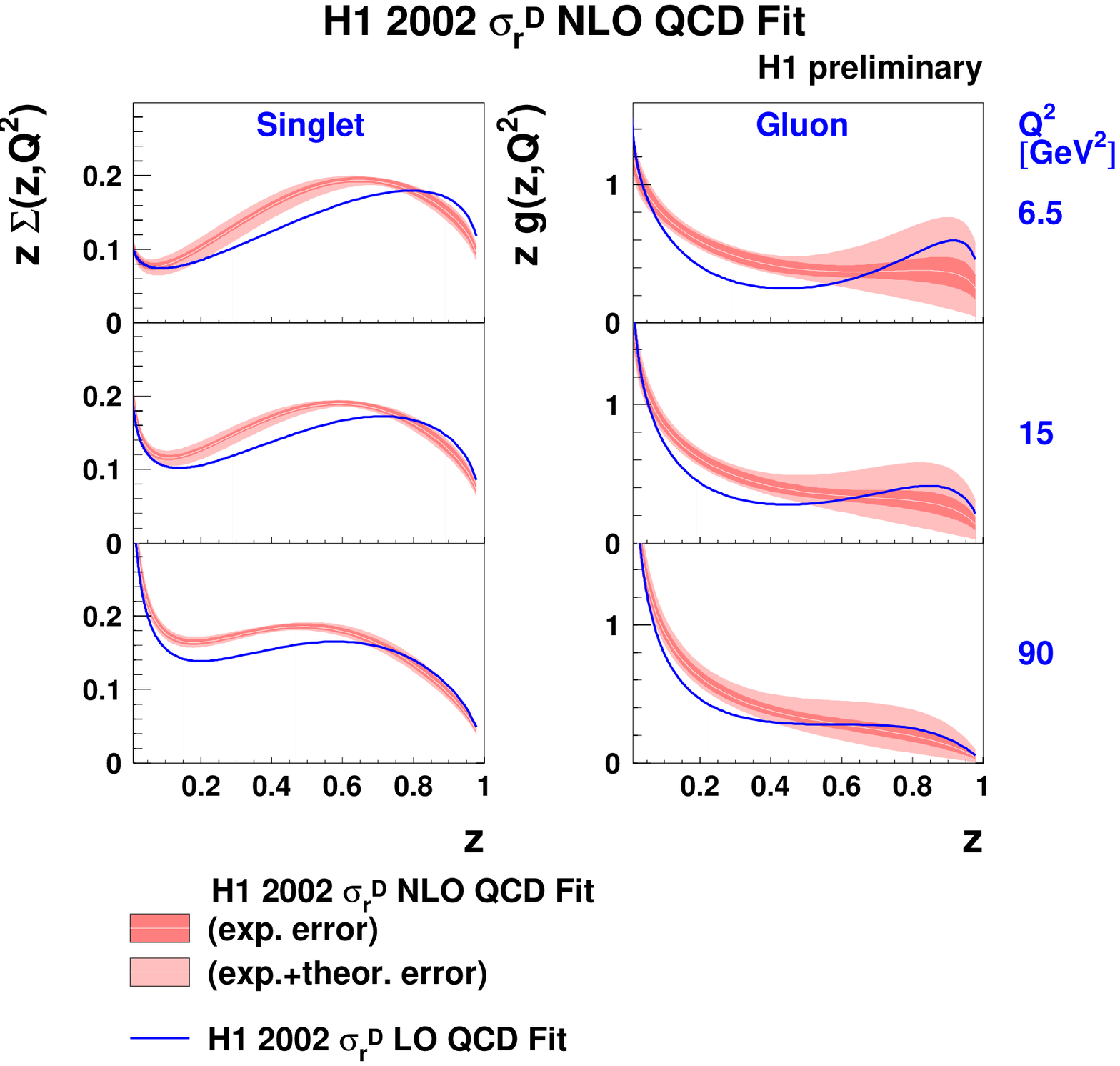,width=0.9\textwidth}
 %\vspace{-0.5cm}
 \caption{
Quark (singlet) and gluon densities in the $\pom$ extracted from the QCD 
fit of \fdthree as a function of $z$.}
\label{fig:gluond}
\end{figure}

The reggeon contribution gets larger for increasing values of
$\xpom$, which correspond to smaller energy (for given \qsq\ and \bet\
values).
It also gets larger for smaller values of \bet, which is consistent with
the expected decrease with \bet\ of the reggeon structure function, following
the meson example, whereas the Pomeron structure function is observed to
be approximately flat in \bet.
The comparision of \fdtwo\ with the inclusive proton structure function,
\ftwo,
shows that the dynamic is different in the Pomeron and in the proton 
(see Fig.~\ref{fig:f2d}) for large \bet\ values.  
 By analogy to the QCD evolution of the proton structure function, 
one can attempt to extract the partonic structure of the Pomeron
from the \qsq \ evolution of \fdtwo. 
The extracted distributions\cite{f2d} are shown in Fig.~\ref{fig:gluond}
separately for the gluon 
and the singlet quark components as a function of $z$, the Pomeron
momentum fraction carried by the parton entering the hard interaction.
This distribution shows the dominance of gluons up to the high $z$ values
in the Pomeron partonic structure.
\\

 The dominance of hard gluons in the Pomeron is confirmed by various
analyses of the diffractive hadronic final 
state (jet production, energy flow, particle spectra and multiplicities,
and event shape) providing a global consistent picture of 
diffraction\cite{hadfin,mult,jets}.
%%%%%%%%%%%%%%%%%%%%%%%%%%%%%%%%%%%%%%%%%%%%%%%%%%%%%%%%%%%%%%%%%%%%%%%%%

\section{Exclusive diffractive final states}
%%%%%%%%%%%%%%%%%%%%%%%%%%%%%%%%%%%%%%%%%%%
\begin{figure}[htb] \unitlength 1mm
 \begin{center}
  \epsfig{file=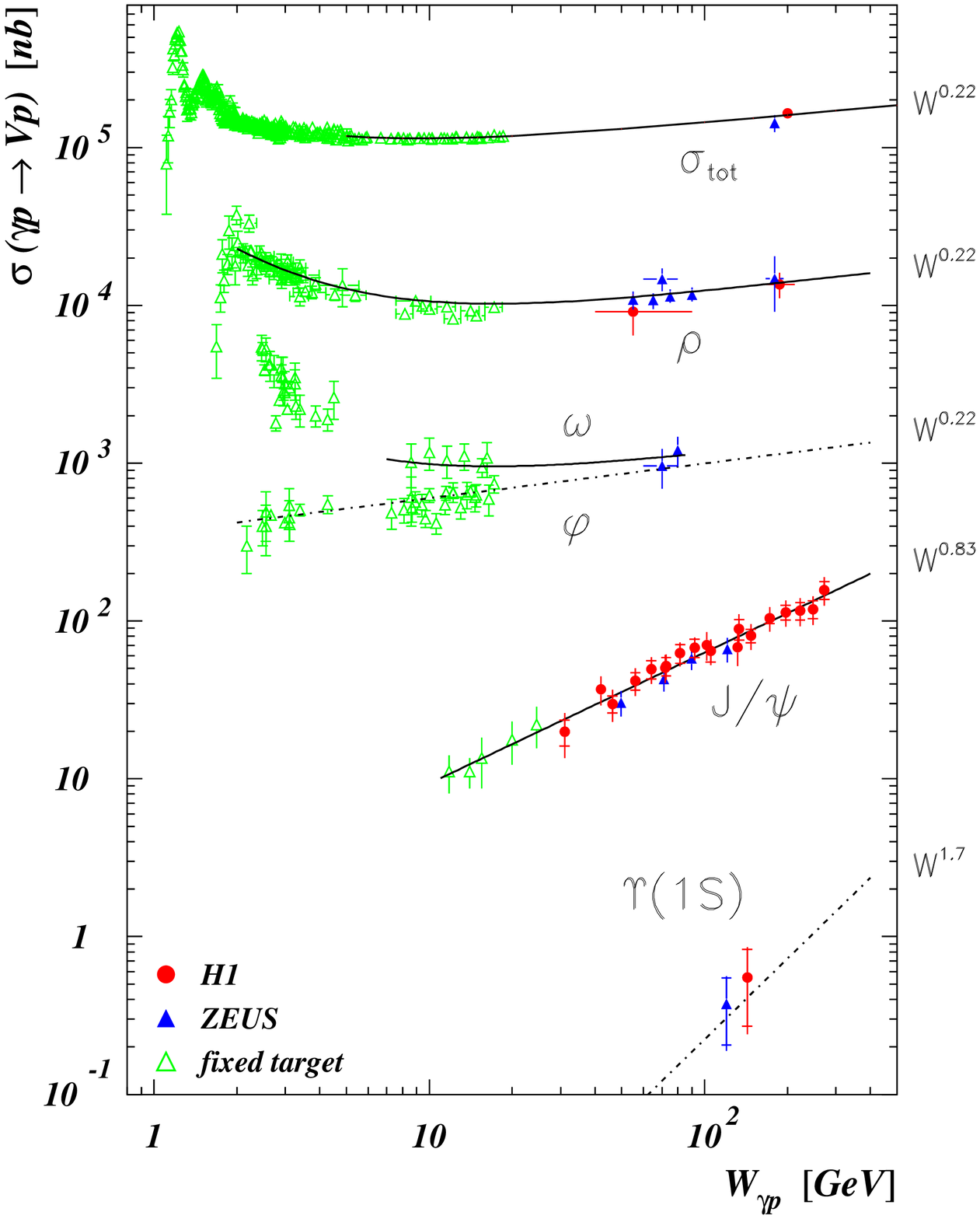,width=0.85\textwidth}
 \end{center}
 \caption{Diffractive cross sections as a function of the $\gamma^*-p$
system energy for various vector mesons in photoproduction together 
with the total photoproduction cross section.
}
 \vspace{-0.5cm}
\label{fig:vm}
\end{figure}

 During these last
decades, our knowledge of the hadronic structure has progressed a lot,  
mostly with the study of the inclusive inelastic reactions. The study of 
exclusive reactions will probably be dominant in the coming years, in
this field, as they allow us to go one step deeper in our understanding
of the hadronic structure. In particular they give access to
momenta and helicity correlations of the hadron structure. 
\\

 Furthermore, the exclusive elastic vector meson production and the 
deeply virtual Compton scattering (DVCS) studies: 
$\gamma^* p \rightarrow X p$, where the hadronic system $X$ consists 
only in a vector meson ($\rho, \omega,...$) or a photon (DVCS), 
provide a very 
interesting way to test the mechanism of diffraction and our 
understanding of the Pomeron structure.
\\

Fig.~\ref{fig:vm} summarizes the $W$ dependence of various 
elastic exclusive vector meson photo-production\cite{vm}, 
together with the total photoproduction cross section\cite{totsig}. 
The light mesons ($\rho, \omega$ and $\phi$) show a soft dependence in $W$, 
equivalent to that of the total cross section dependence, 
while this energy dependence is much steeper
for $J/\psi$ production.
This regime change is interpreted as being due to the presence of a hard scale
in the process, the (high) charm 
quark mass, making the $J/\psi$ meson smaller than the
confinement scale ($\sim 1 fm$). 
\stepcounter{figure}
\setcounter{fig:jpsi}{\value{figure}}
In this case, it is natural to attempt a perturbative QCD description
of the process,
where the photon fluctuates into a quark-antiquark pair
and the exchanged Pomeron is modeled by a pair of gluons.
This leads, in a first approximation, to a cross section proportional 
to the gluon density squared, which is in good agreement
(full line) with the data (points) shown on 
Fig.~\arabic{fig:jpsi}.
\stepcounter{figure}
\setcounter{fig:rho}{\value{figure}}

\hspace{-0.8cm}
\begin{minipage}[h]{0.48\textwidth}
 \vspace{0.4cm}
 \epsfig{file=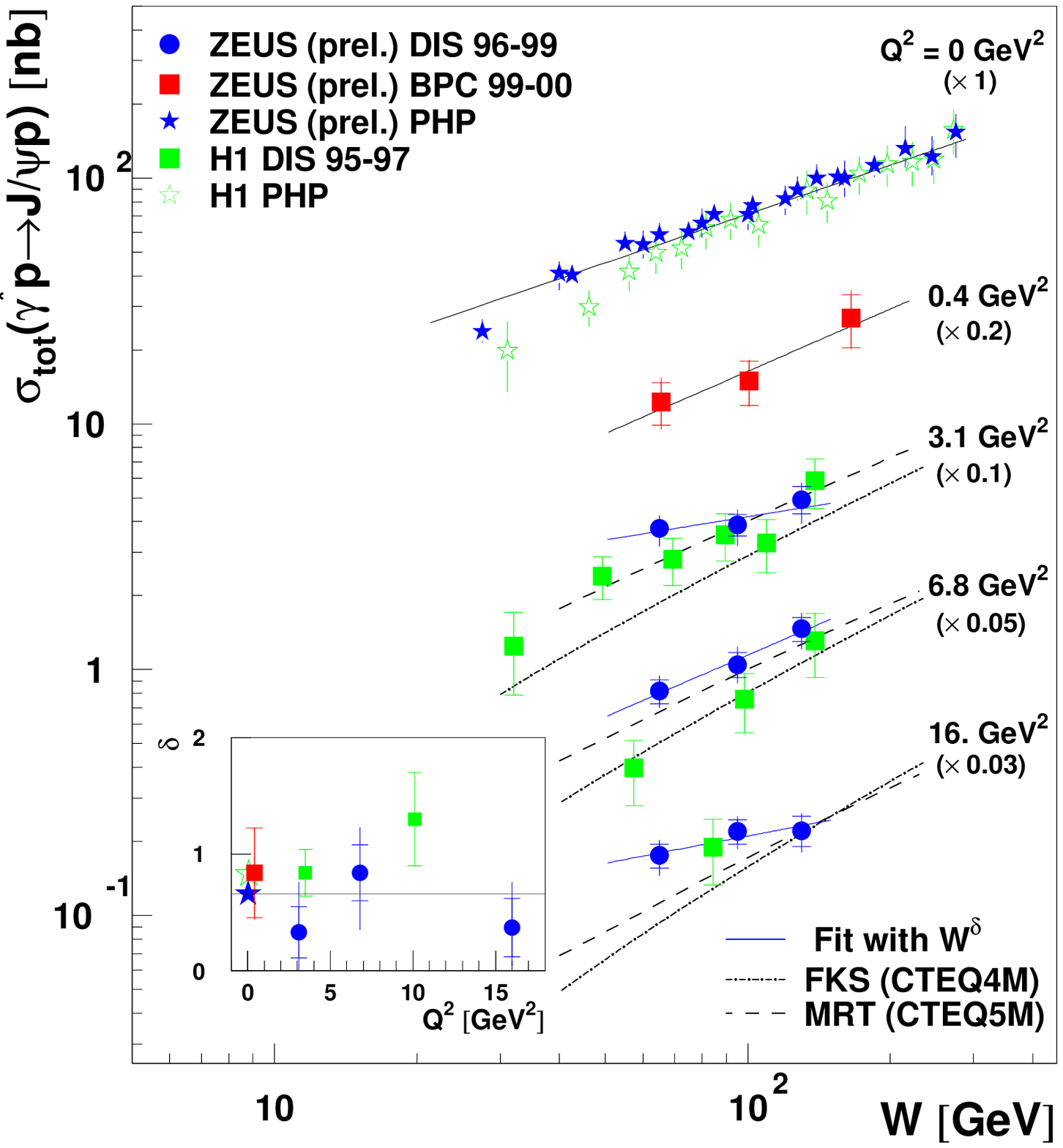,height=6.0cm}
 \small{Figure~\arabic{fig:jpsi}. Exclusive $J/\psi$ cross section in 
  photoproduction and DIS Diffractive cross sections as a function of $W$.}
  \\
\end{minipage}
\hspace{0.3cm}
\begin{minipage}[h]{0.48\textwidth}
 \epsfig{file=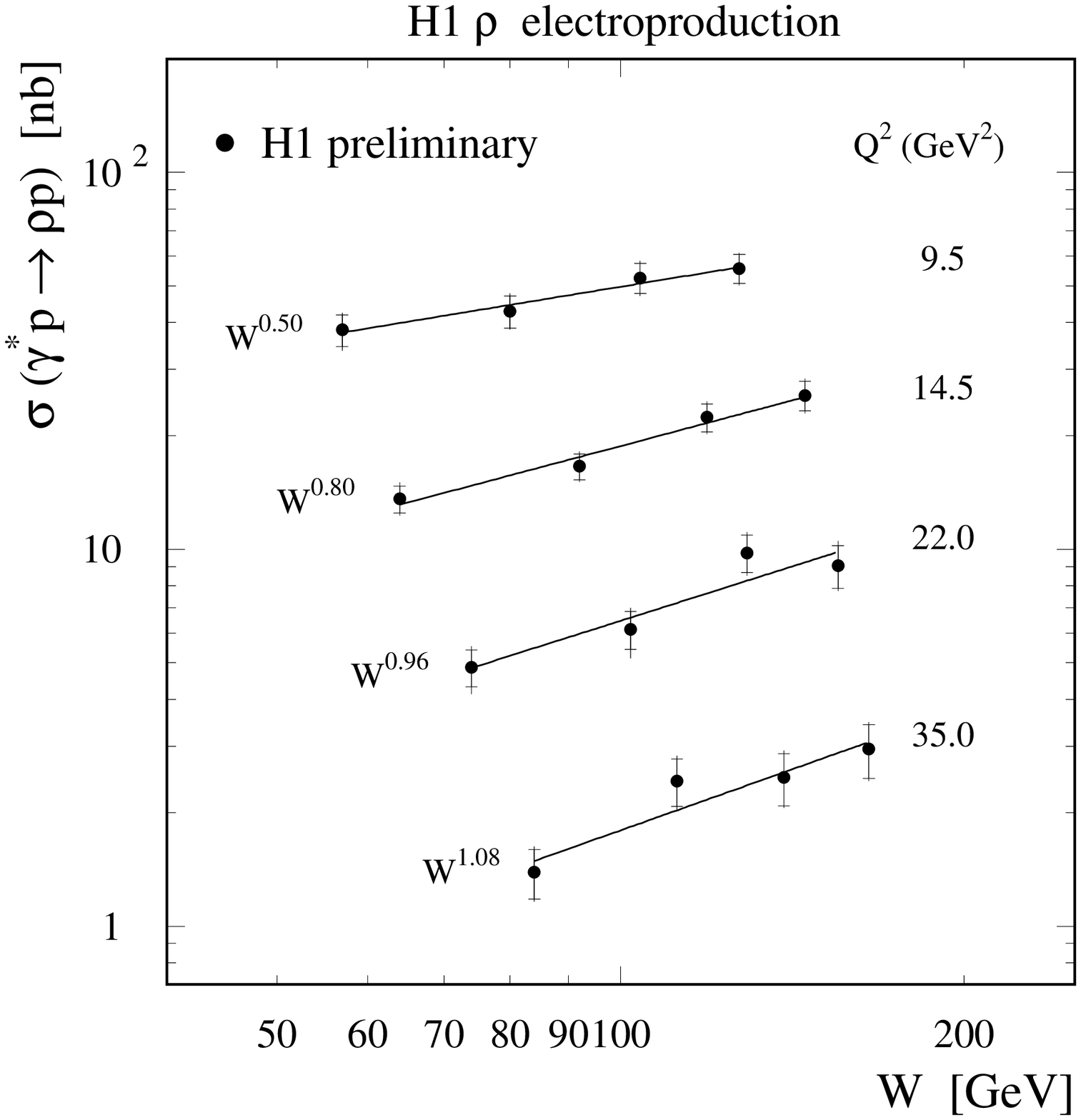,height=6.0cm}
 \small{Figure~\arabic{fig:rho}. Exclusive $\rho$ in DIS for different 
  $Q^2$ values.}
 \\
\end{minipage}

This figure also shows the agreement of the 2 gluons exchange model
with measurement of exclusive  $J/\psi$ production in the DIS regime, 
where a second hard scale, \qsq\ is present.
As illustrated on Fig.~\arabic{fig:rho}, a modification of the $W$ 
dependence also occurs for the elastic $\rho$ production
when the $Q^2$ increases, which highlights the appearance of \qsq\ 
as a hard scale for a light vector meson production. 
\\

\begin{figure}[btp]
 \setlength{\unitlength}{1.0cm}
 \begin{picture}(12.0,8.0)
  \put(2.0,0.){\epsfig{file=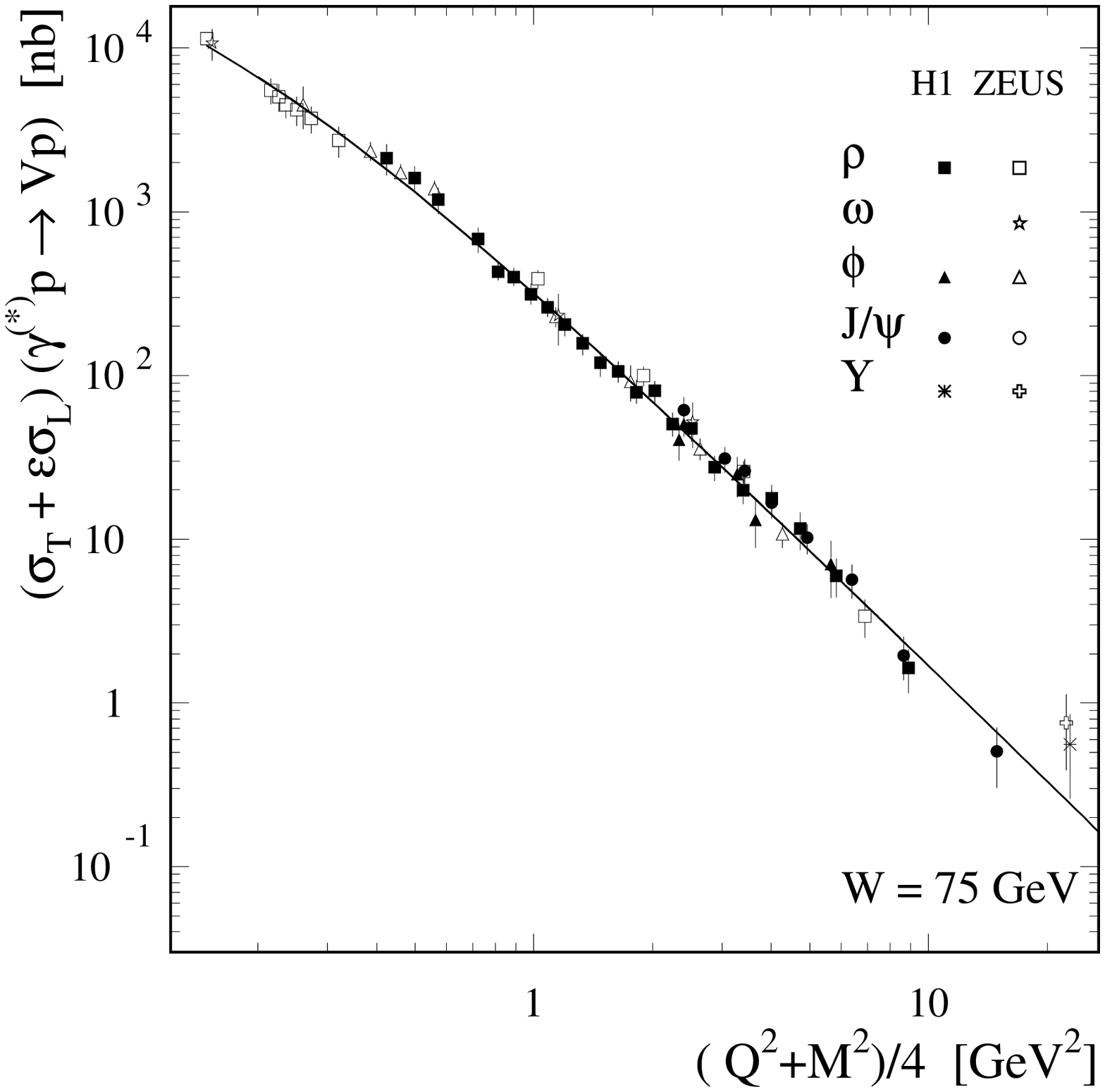,width=0.75\textwidth}}
  \put(3.5,1.3){\epsfig{file=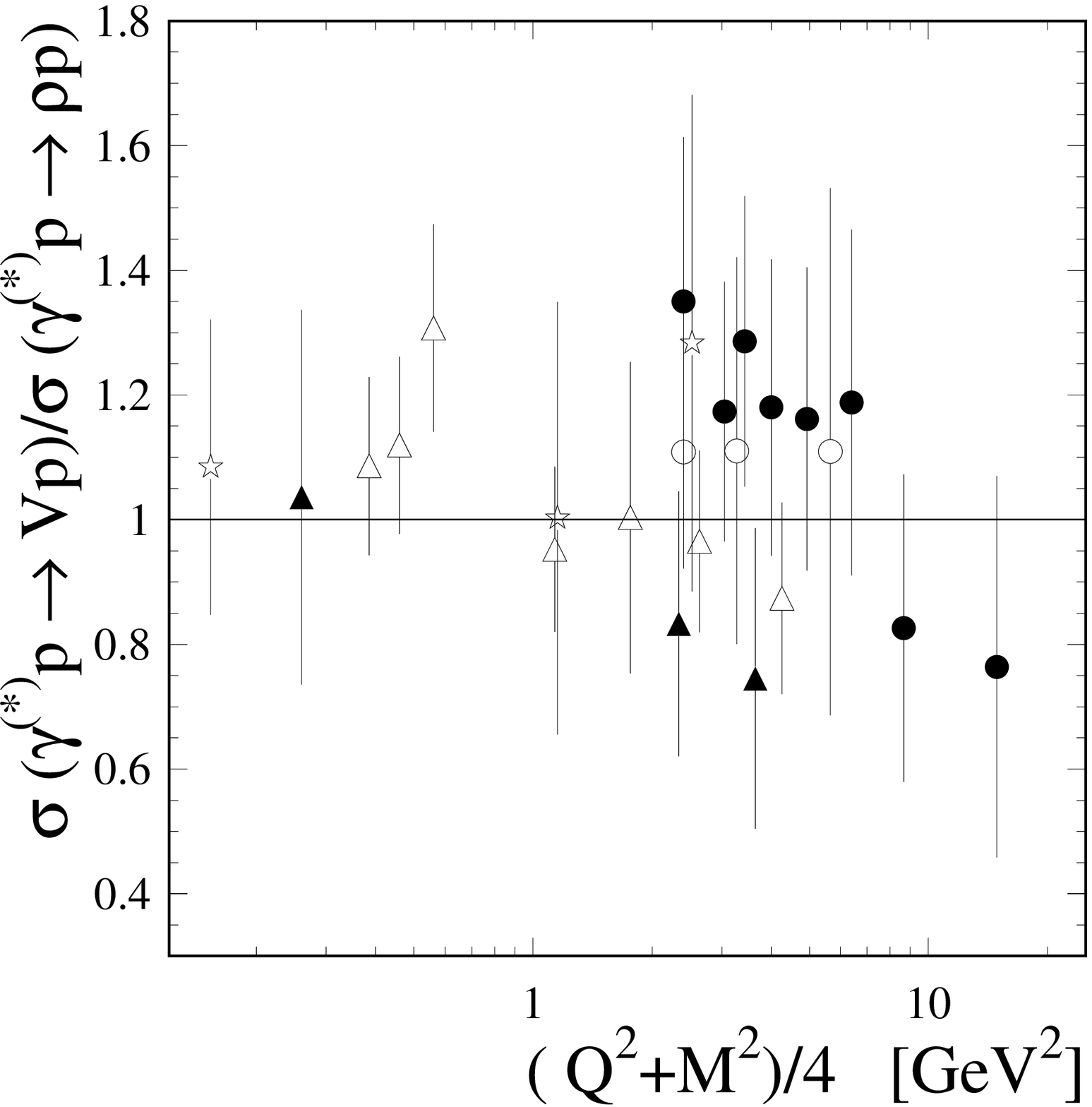,width=3.5cm}}
 \end{picture}
 \caption{H1 and ZEUS measurements of the $\gamma p \rightarrow VM p$
  total cross sections as a function of (\qsq\ + $M^2_V$)/4
  for $\rho, \omega, \phi, J/\psi$ and $\Upsilon$ vector meson elastic
  production, at the fixed value of $W$ = 75 GeV.
  The insert presents the distance of the points to the fitted curve.
  }
 \label{fig:univers}
\end{figure}

 It is remarkable to see that the different vector meson
cross sections present a universal behaviour as a function of the
variable $Q^2 + M_V^2$, where $M_V$ is the vector meson mass, as
shown in Fig.~\ref{fig:univers} for  $W$ = 75 GeV.
The cross sections were scaled by SU(4) factors, according to the
quark content of the vector mesons\cite{vm}. 
This indicates that this variable choice seems to be a good scale to
describe the vector meson production in a universal way, and that the
cross section is at the first order not sensitive to the quark
flavour (which would probably not be valid at lower energy).
\\

 However, the exclusive $\Upsilon$ cross section seems to be too high 
for this universal behaviour. In fact, the cross section measurement
does not agree with simple QCD predictions.
This is due to the fact that 
the transition from a virtual photon coming from an electron 
to an on shell particle forces the fractional momenta of the two 
gluons involved to be unequal.
Therefore, in the cross section expression, the gluon density cannot just be
squared, but two different $x$ values have to be considered. 
When such skewing effects are included the agreement with
QCD can be found again.
This effect becomes important at high \qsq\ values or high vector meson
masses. To include such effects in the cross section calculations, the
very powerfull formalism of Generalised Parton Distributions (GPD) has 
been introduced\cite{GPD}. 
\\

\begin{figure}[btp]
 \setlength{\unitlength}{1.0cm}
 \begin{picture}(12.0,6.0)
  \put(-0.2,0.){\epsfig{file=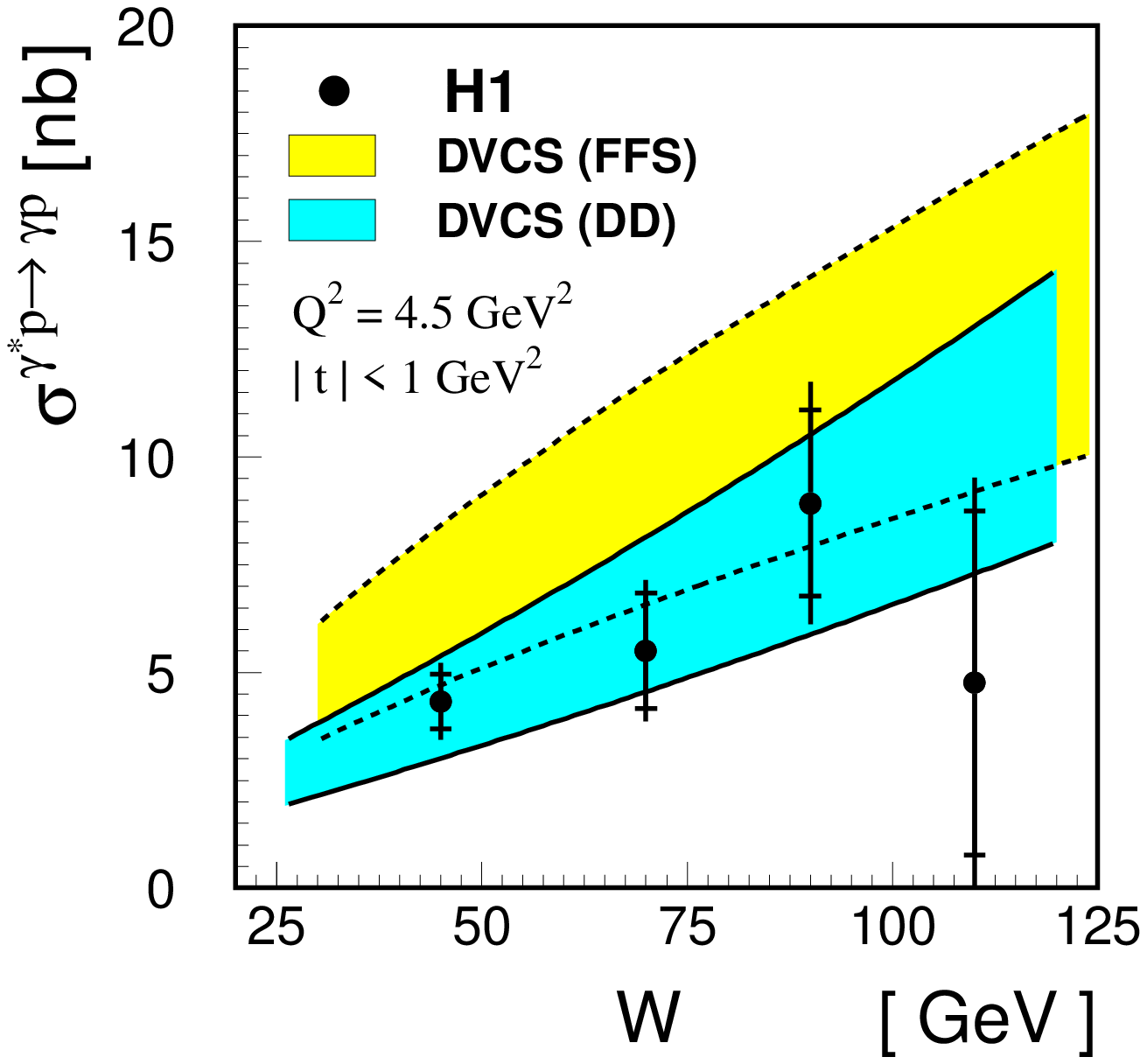,width=0.45\textwidth}}
  \put(5.5,0.25){\epsfig{file=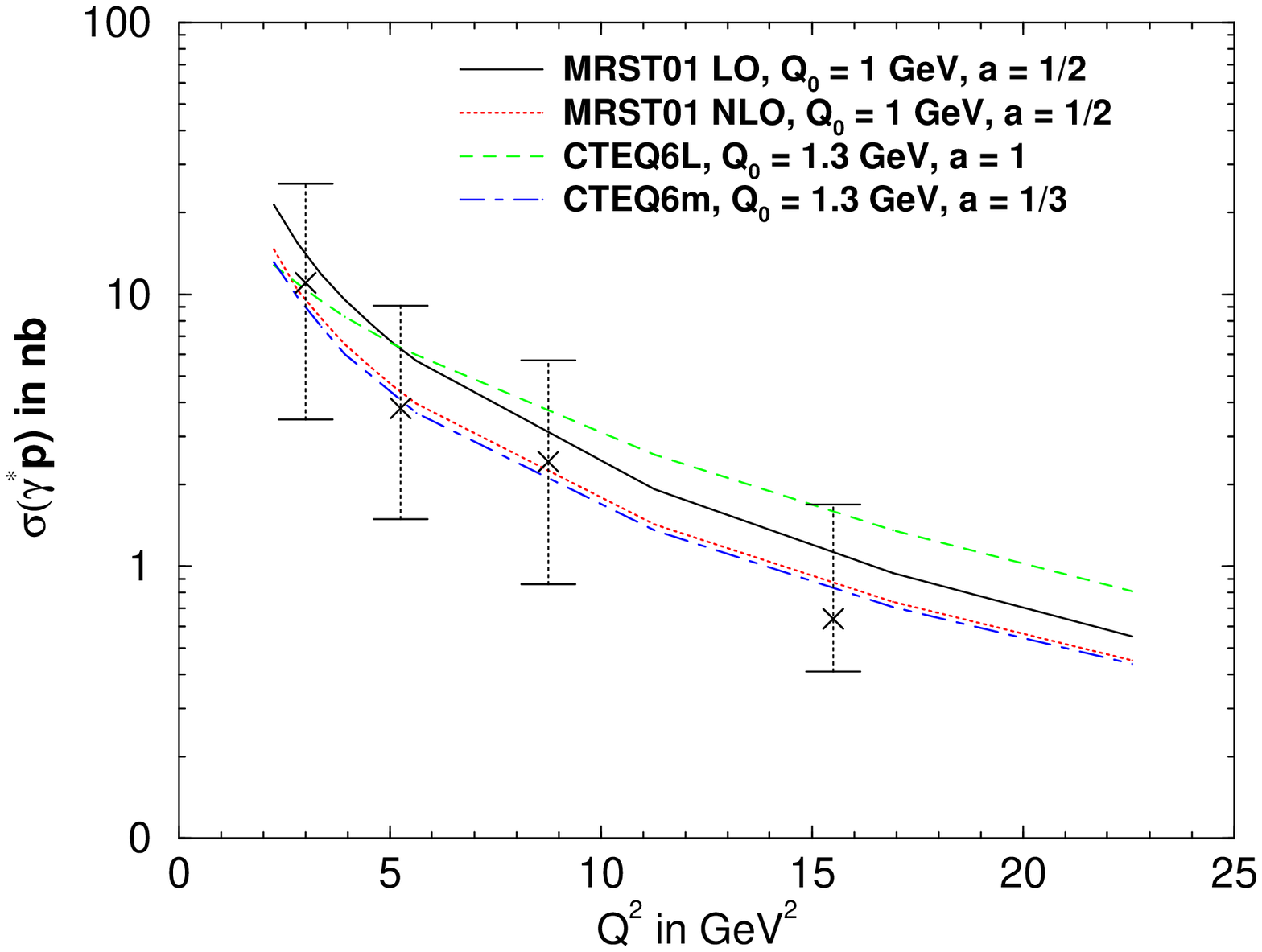,width=0.53\textwidth}}
 \end{picture}
 \caption{The H1 DVCS cross section measurement as a function of $W$
 (left) and of \qsq\ (right). The measurement is compared to 
 dipole model (DD) and QCD predictions (FFS for LO and right plot at 
 NLO with different GPD inputs).
  }
 \label{fig:dvcs}
\end{figure}
For vector meson production GPD appear
quadratically in the cross section expression.
A unique feature of DVCS is that they appear linearly in the
interference term with the purely QED Bethe-Heitler process.
Compared to vector meson production, DVCS is theoretically simpler
(fully calculable)
because the composite meson in the final state is replaced by
the photon, thus avoiding large uncertainties due to the
unknown meson wave functions.
Therefore the DVCS measurement
offers a particularly suitable channel to extract GPD. 
The first cross section measurement\cite{dvcs}, shown in Fig.~\ref{fig:dvcs} 
as a function of $W$ and $Q^2$, is in good agreement with the
perturbative QCD prediction\cite{Freund} and provides the first
constrains on GPDs\cite{Ferrare}.

%%%%%%%%%%%%%%%%%%%%%%%%%%%%%%%%%%%%%%%%%%%%%%%%%%%%%%%%%%%%%%%%%%%%%%%%%
\section*{Conclusion}
%%%%%%%%%%%%%%%%%%%%%

HERA experiments have produced a large amount of results in diffraction,
which allow confrontations with QCD predictions,
when one of the hard scales \qsq, the quark mass or \ttt \ 
(not reported in this summary) is present in the process.
\\

 The QCD analysis of the total diffractive cross section, 
assuming factorisation into a Pomeron flux in the proton, 
shows that the corresponding parton distributions 
 favors the dominance of hard gluons in the Pomeron. 
This is confirmed by the analysis
 of inclusive final states and of jet production.
\\

 For the case of exclusive final state production, in the presence of a 
hard scale the transition from soft to hard regime is observed as in 
inclusive DIS.
Models based on the fluctuation of the photon in a quark-antiquark pair which
subsequently exchange a pair of gluons with the proton parton successfully 
reproduce the enhanced energy dependence for the $J/\psi$ production and
for the $\Upsilon$ when skewed effects are included. The DVCS cross
section, the only fully calculable diffractive process, 
has been measured for the first time and is in good agreement
with QCD predictions.

\end{document}